\documentclass{amsart}
\usepackage{amssymb}
\usepackage{amsmath}
\usepackage{amsfonts}
\usepackage{graphicx}
\usepackage{epstopdf}
\usepackage{hyperref}

\newcommand{\beq}{\begin{equation}}
\newcommand{\eeq}{\end{equation}}

\begin{document}

\begin{center}
{\Large \bf Universal Chern-Simons partition functions  as  quadruple Barnes' gamma-functions} \\
\vspace*{1 cm}
{\large  R.L.Mkrtchyan 
}
\vspace*{0.5 cm}

{\small\it Yerevan Physics Institute, 2 Alikhanian Br. Str., 0036 Yerevan, Armenia}

\end{center}\vspace{2cm}

{\small  {\bf Abstract.} We show that both perturbative and non-perturbative parts of universal partition functions of Chern-Simons theory on 3d sphere are ratios of four over four Barnes' quadruple gamma functions with arguments given by linear combinations of universal parameters. Since nonperturbative part of partition function is essentially a universal compact simple Lie group's volume, latter appears to be expressed through quadruple Barnes' functions, also. For SU(N) values of parameters recurrent relations on Barnes' functions give the proof of level-rank duality of complete partition function, thus extending that duality on non-integer level and rank. We note that integral representation of universal partition function is defined on few disjoint regions in parameters' space, corresponding to different signs of real parts of parameters, and introduce a framework for discussion of analytic continuation of partition functions(s) from these regions. Although initial integral representation is symmetric under all permutations of parameters (which corresponds particularly to $N \rightarrow -N$ duality of gauge theories with classical groups), analytic continuations are not symmetric under transposition of parameters with different signs of their real parts. For the  particular case of SU(N) Chern-Simons this asymmetry appears to be the Kinkelin's functional equation (reflection relation) for Barnes' G-function. }

%\tableofcontents

\section{Introduction}\label{intro}

Recently a closed explicit formula through universal parameters have been obtained for partition function $Z$ of Chern-Simons theory on 3d sphere \cite{Mkr3}.  It is a transformation of formula \cite{W1} $Z=S_{00}$ for partition function through matrix element of  modular S-transformation, with simultaneous extension to an arbitrary values of Vogel's universal parameters \cite{V0,V}.  

Partition function is the product of perturbative and nonperturbative parts $Z=Z_1Z_2$. For perturbative free energy we have \cite{Mkr3}:
\begin{eqnarray}\label{F2}
F_2=-\ln Z_2 = \int^{\infty}_0 \frac{dx}{x} \frac{F(x/\delta)}{(e^{x}-1)}
\end{eqnarray}
where $F(x)$ is a universal expression \cite{MV1} for character of adjoint representation $f(x)$, evaluated at point $x \rho$, minus dimension of algebra $dim$.  $\rho$ is a Weyl vector in roots space, i.e. half of sum of all positive roots of a given simple Lie algebra.

\begin{eqnarray}\label{gene}
F(x)&=&f(x)-dim\\
f(x)&=&\frac{\sinh(x\frac{\alpha-2t}{4})}{\sinh(x\frac{\alpha}{4})}\frac{\sinh(x\frac{\beta-2t}{4})}{\sinh(x\frac{\beta}{4})}\frac{\sinh(x\frac{\gamma-2t}{4})}{\sinh(x\frac{\gamma}{4})} \\ \label{dim}
dim &=& \frac{(\alpha-2t)(\beta-2t)(\gamma-2t)}{\alpha\beta\gamma}\\ 
\delta&=&\kappa+t \\
 t&=&\alpha+\beta+\gamma
\end{eqnarray}
$\kappa$ is a coefficient in front of Chern-Simons action, $\delta$ is (an inverse) usual effective coupling constant equal to the sum of bar coupling $\kappa$ and half of eigenvalue $2t$ of second Casimir operator in adjoint representation. $\alpha, \beta, \gamma$ are
universal parameters introduced by Vogel \cite{V0}. They are projective parameters, with correspondence with simple Lie algebras given by Table \ref{tab:1}:

\begin{table}[h] \label{tab:1}
\caption{Vogel's parameters for simple Lie algebras}     
\begin{tabular}{|r|r|r|r|r|} 
\hline Algebra/Parameters & $\alpha$ &$\beta$  &$\gamma$  & t \\ 
\hline SU(n) & -2 & 2 & n & n \\ 
\hline SO(n)/Sp(-n) & -2  & 4 & n-4 & n-2 \\ 
\hline Exc(n) & -2 & 2n+4  & n+4 & 3n+6 \\ 
\hline 
\end{tabular} 
\end{table}
where for SU(n) and SO(n) n is positive integer, for Sp(-n) n is negative even, for exceptional line Exc(n) n=-2/3,0,1,2,4,8 for  $G_2, D_4, F_4, E_6, E_7, E_8$  respectively. For Chern-Simons theory projective is set $\alpha,\beta,\gamma,\delta$. Values of parameters in Table \ref{tab:1} are given in so called minimal normalization, when square of long roots is 2. In this normalization physical values of parameter $\kappa$ are non-negative integers $k$. 

Universal are called quantities, which can be expressed through Vogel's parameters by smooth (analytic, rational, etc.) functions, in such a way that their values at points from Table give an answers for theory with corresponding algebras. Perturbative partition function (\ref{F2}) is an example of universal quantity, as well as unknot Wilson loop, dimension of simple Lie algebra (\ref{dim}) (and some classes of their irreps \cite{LM1}), central charge, etc. \cite{MV1}. 

Non-perturbative part of partition function is expressed through perturbative one by observation \cite{Mkr3} that $ Z=1 $ at $\kappa=0$. This follows from $S_{00}=1$ at $\kappa=0$, since there is no contributions into $S_{00}$, i.e. there is no integrable representation of corresponding affine Kac-Moody algebra at level k=0, besides trivial one \cite{W1}. 

From this one deduces nonperturbative free energy (see\cite{Mkr3}): 

\begin{eqnarray}
F_1=-\ln Z_1= (dim/2)\ln(\delta/t) -\int^{\infty}_0 \frac{dx}{x} \frac{F(x/t)}{(e^{x}-1)}\label{F1}
 \end{eqnarray}

and complete free energy:
\begin{eqnarray}\label{totalfree}
F_1+F_2  =(dim/2)\ln(\delta/t)+
\int^{\infty}_0 \frac{dx}{x} \frac{F(x/\delta)-F(x/t)}{(e^{x}-1)}=\\
(dim/2)\ln(\delta/t)+
\int^{\infty}_0 \frac{dx}{x} \frac{f(x/\delta)-f(x/t)}{(e^{x}-1)} \label{Ftotal}
 \end{eqnarray}
Note that in last integral terms with $dim$ canceled. 

Finally, universal formula for volume (as defined in \cite{MV1}) of corresponding compact simple Lie group with Cartan-Killing metric is \cite{Mkr3}:

\begin{eqnarray}
Vol(G)&=& (2\pi t^{-1/2})^{dim}Z_2|_{\kappa=0}\\
&=&(2\pi t^{-1/2})^{dim}\exp{\left( -\int^{\infty}_0 \frac{dx}{x} \frac{F(x/t)}{(e^{x}-1)}\right) } \label{vol}
\end{eqnarray}

The main aim of present paper is to make contact with the field of number theory functions, particularly Barnes' multiple gamma-functions \cite{Barnes3}. These functions appear often in integrable theories, see e.g. recent discussion of double gamma-function in \cite{Fad}. We show that both perturbative and non-perturbative partition functions are combinations of Barnes' quadruple gamma-functions with arguments and parameters linearly depending on universal parameters  $\alpha,\beta,\gamma,\delta$.

The basic object is Barnes' multiple zeta-function \cite{Barnes3}, generalization of Riemann and Hurwitz zeta functions. It is given by 
 
\begin{eqnarray}\label{zN}
\zeta_N(w,s|a_1,a_2,...,a_N)=\sum_{n_1,...,n_N=0}^{\infty}\frac{1}{(w+a_1n_1+a_2n_2+...+a_N n_N)^s}
\end{eqnarray}
defined particularly at $\Re w >0, \Re s>N $ and positive parameters $a_i$.

Barnes's multiple gamma-functions $\Gamma_N(w)$ are in  similar connection with Barnes' multiple zeta functions as usual gamma-function with Riemann's zeta-function: 
 
\begin{eqnarray}\label{gN}
\ln\Gamma_N(w)=\Psi_N(w)=\Psi_N(w|a_1,a_2,...,a_N)=\\
\partial\zeta_N(w,s|a_1,a_2,...,a_N)|_{s=0}
\end{eqnarray}
This definition follows that of \cite{Rui} and differs from original Barnes' one \cite{Barnes3} by some modular "constant", depending on parameters. It seems to be more convenient for our purposes.

With these definitions perturbative free energy is equal to (Section \ref{barnes})
\begin{eqnarray}\label{pfgn}
Z_2= \frac{\Gamma_4(w_1) \Gamma_4(w_2) \Gamma_4(w_3) \Gamma_4(w_7)  }{\Gamma_4(w_4) \Gamma_4(w_5) \Gamma_4(w_6) \Gamma_4(w_8) }\left( \frac{\delta}{\pi} \right)^{\frac{dim}{2}}
\end{eqnarray}
where arguments $w_i, i=1,2,...,8$ of quadruple gamma functions are given by (\ref{ws}), parameters are $ -\alpha,\beta,\gamma,2\delta$. 

The non-perturbative part of partition function is expressed through quadruple Barnes' function essentially by above formula, with $\kappa=0$, see (\ref{npfgn}). 

This last expression, due to (\ref{vol}), give rise to formula for invariant volume of compact simple Lie groups through Barnes' quadruple gamma function. This is  beautiful and unexpected relation between number theory functions and intrinsic geometrical properties of simple Lie algebras and groups.

Complete partition function appears to be expressed through Barnes' quadruple gamma functions, and this representation  leads to a simple proof of level-rank duality \cite{NS,CLZ}for $SU(N)$ partition function in Section \ref{levelrank}. The proof is based on the recurrent relations of multiple gamma functions. This  extends the level-rank duality on non-integer values of $N$ and $k$, whereas existing proofs usually are dealing with integer $N$ and $k$, only.

In Section \ref{reflect} we discuss analytic properties of partition function. We argue that it is equal to different analytic functions in different regions of parameter's space (differing by signs of real parts of parameters). The reason  is  noted in \cite{Mkr3} on the example of toy model

\begin{eqnarray}\label{toy1}
\int_{0}^{\infty}\frac{dx}{\cosh(zx)}= \left\{ \begin{array}{rl}
 \frac{\pi}{2z} &\mbox{if $\Re z >0$} \\
 -\frac{\pi}{2z}&\mbox{if $\Re z <0$}
       \end{array} \right.
\end{eqnarray}
or one can consider even simpler integral

\begin{eqnarray} \label{toy2}
\int_{0}^{\infty}\frac{dx}{1+(zx)^2}  =   \left\{ \begin{array}{rl}
 \frac{\pi}{2z} &\mbox{if $\Re z >0$} \\
 -\frac{\pi}{2z}&\mbox{if $\Re z <0$}
       \end{array} \right.
\end{eqnarray}

They are  symmetric under change of sign of $z$, and  define two analytic functions, connected by the change of sign of $ z$.
These are analytic functions $f_+(z)=\pi/2z$ or $f_-(z)=-\pi/2z$, for positive and negative real parts of arguments, respectively. The reason is that one cannot connect  two points on a complex plane of parameter $z$, one with positive real part and another one with negative, by a path without passing through singularity of integrals (\ref{toy2})(\ref{toy1}). Namely, one can't avoid crossing the line $\Re z=0$, any point on which is singular for integrals. Values of integrals at $\Re z>0$ and  $\Re z<0$ doesn't  belong to the same analytic function, instead they are given by two different analytic functions $f_+(z)$ and $f_-(z)$.   Each of these functions is initially defined in the corresponding region of convergence of integral (\ref{toy2})(\ref{toy1}), i.e. corresponding open semiplane, but then it can be analytically continued to another half, where it can be compared with function originated from that half. Thus, first, they are connected by $z \rightarrow -z$ transformation, and, second, their difference is $f_+(z)-f_-(z)=\pi/z$.

Similarly, as discussed in Section \ref{reflect}, expression \ref{F2} defines few (sets of) analytic functions, corresponding to different sets of real parts of parameters.  Initial function $F_2$ is symmetric w.r.t. the permutations of parameters, but these analytic functions  already not, rather functions in  the same set are connected by permutations of parameters. One can be interested in calculation of their change under that permutations. Such a relations are known in many particular cases and are called reflection relations, functional equations, duality relations, etc. See e.g. \cite{Rui2} for a study of  their role in integrable systems.  Perhaps most famous example of reflection relation is  Euler's  formula for gamma function:

\begin{eqnarray}
\Gamma(1+z)\Gamma(1-z)=\frac{\pi z}{\sin(\pi z)}
\end{eqnarray}

For our toy model reflection relation is calculated above. For  $F_2$ function in particular case of parameters for $SU(N)$ group corresponding reflection relation is essentially the well-known Kinkelin's  functional equation \cite{Kin} for Barnes' G-function \cite{Barnes1}:

\begin{eqnarray} \label{Kin0}
\ln \frac{G(1+N)}{G(1-N)}= N\ln(2\pi )- \int_{0}^{N}dx \, \pi x \, cot(\pi x)
\end{eqnarray}
We derive it from our formulae in Section \ref{reflect} and discuss there the general case, for which reflection relations aren't derived yet. 

In Conclusion we discuss results and possible directions of development.

\section{ Chern-Simons' universal partition functions as Barnes' quadruple gamma functions}\label{barnes}

Barnes' multiple zeta function (\ref{zN}) can be written as an integral

\begin{eqnarray}\label{zNi}
\zeta_N(w,s|a_1,a_2,...,a_N)=\frac{1}{\Gamma(s)}\int \frac{dt}{t}t^s e^{-wt}\prod_{k=1}^{N}\frac{1}{(1-e^{-a_kt})}
\end{eqnarray}

This gives  (\ref{zN}), if one expand all multipliers as $(1-e^{-a_kt})^{-1}=\sum_{n=1}^{\infty}e^{-na_kt}$, multiply the sums and integrate term by term. With this integral representation of zeta function Barnes' multiple (N-ple) gamma-function (\ref{gN}) can be represented  \cite{Rui} as:

\begin{eqnarray} \label{Psi}
\Psi_N(w)=\int_{0}^{\infty}\frac{dx}{x}\left( e^{-wx} \prod_{j=1}^{N} \frac{1}{(1-e^{-a_jx})} \right.  \\
\left. - x^  {-N}\sum_{n=0}^{N-1}\frac{(-x^n)}{n!}B_{N,n}(w) -\frac{(-1)^N}{N!}e^{-x}B_{N,N}(w)\right) 
\end{eqnarray}
where multiple Bernoulli  polynomials $B_{N,n}(w)$ are defined as:
\begin{eqnarray}
x^{N}e^{-wx}\prod_{j=1}^{N} \frac{1}{(1-e^{-a_jx})}= \sum_{n=0}^{\infty}\frac{(-x^n)}{n!}B_{N,n}(w)
\end{eqnarray}

Both gamma function and multiple Bernoulli polynomials are implied to depend on positive parameters $ a_j $.  Integrals converge provided real part of $w$ is positive. 

With these definitions let's express perturbative partition function through quadruple  Barnes' gamma functions.

Assume parameters are in the physical region $\alpha<0,\beta>0,\gamma>0, t>0,\kappa>0$. Then, separating truly negative exponents in denominators and expanding nominator into the sum of exponents, we can present $F_2$ as:

  \begin{eqnarray}
F_2=\int \frac{dx}{x}  \left(   \frac{-\sum_{i=1}^{8}\epsilon_i e^{-w_i x}}{(1-e^{-2\delta x})(1-e^{\alpha x})(1-e^{-\beta x})(1-e^{-\gamma x})} \right.   \\
 \left.  -dim \frac{ e^{-2\delta x}}{(1-e^{-2\delta x})}  \right)  
    \end{eqnarray}
where $\epsilon_i=+,+,+,-,-,-,+,-$, and $w$-s are:

\begin{eqnarray}\label{ws}
w_1&=& 2\delta-2\alpha,\\
w_2&=& 2\delta-\alpha-\beta,\\
w_3&=& 2\delta-\alpha-\gamma,\\
w_4&=& 2\delta+\alpha+\beta+\gamma,\\
w_5&=& 2\delta+2\beta+\gamma,\\
w_6&=& 2\delta+\beta+2\gamma,\\
w_7&=& 2\delta+2\alpha+3\beta+3\gamma,    \\ \label{ws2}
w_8&=& 2\delta-3\alpha-2\beta-2\gamma,
\end{eqnarray}

This expression of $F_2$ is a combination, with coefficients $\epsilon_i$, of first terms in integral representation (\ref{Psi}) of quadruple and uniple gamma-functions, taken with arguments  (\ref{ws})-(\ref{ws2}) and parameters $(-\alpha,\beta,\gamma,2\delta)$ for quadruple and argument $2\delta$ and parameter $2\delta$ for uniple function. However, subsequent terms from (\ref{Psi}) are completely absent. One can understand that they actually completely cancel. 

Indeed, these two sets of integrals exist separately, i.e. from one side $F_2$, which is a combination of pure first terms (call them also a main terms), and from other side the same combination of main terms with corresponding subsequent terms. Then their difference  written under common integral sign is equal to the integral of combination of these subsequent terms, only, because main terms cancel. But any such combination  can be convergent only if it sum up to zero, otherwise it will diverge at $x=0$. In other words, these subsequent terms are aimed to cancel singularities of main terms at $x=0$, if these singularities already cancel between main terms, their subsequent terms should cancel, also.

We check this cancellation explicitly, also, to avoid any unexpected subtleties. 

So, final answer for $F_2$ through quadruple Barnes' gamma-functions is:
\begin{eqnarray}\label{F2P}
F_2=-\Psi_4(2\delta-2\alpha)-\Psi_4(2\delta-\alpha-\beta)-\\ 
\Psi_4(2\delta-\alpha-\gamma)+\Psi_4(2\delta+\alpha+\beta+\gamma)\\ +\Psi_4(2\delta+2\beta+\gamma)+
\Psi_4(2\delta+\beta+2\gamma)\\-\Psi_4(2\delta+2\alpha+3\beta+3\gamma)+\Psi_4(2\delta-3\alpha-2\beta-2\gamma)\\
- dim \Psi_1(2\delta)
\end{eqnarray}

Implicit parameters in quadruple gamma-functions are $-\alpha,\beta,\gamma,2\delta$, and  implicit parameter in uniple function is $2\delta$. They all are positive, as should be. One can transform in a similar way total free energy in a form (\ref{Ftotal}), then one will get sixteen quintuple Barnes' functions (ratio of eight over eight) and  terms proportional to $dim$ will disappear.

Last term can be simplified. According to \cite{Rui},
\begin{eqnarray}
\Gamma_1(w|a)=exp\left( \left( \frac{w}{a}-\frac{1}{2}\right) \ln a \right) \Gamma\left( \frac{w}{a}\right) (2\pi)^{-\frac{1}{2}}
\end{eqnarray}
where $\Gamma$ in the r.h.s. is usual (Euler's) gamma function. For $w=a=2\delta$, taking into account $\Gamma(1)=1$, we have 
\begin{eqnarray} \label{gamma1}
\Psi_1(2\delta|2\delta)=\ln \Gamma_1(2\delta|2\delta)=\frac{1}{2}\ln\frac{\delta}{\pi}
\end{eqnarray}

Then we get expression (\ref{pfgn}) for perturbative partition function. For nonperturbative one we get from (\ref{F1}): 
\begin{eqnarray}\label{npfgn}
Z_1= \frac{\Gamma_4(v_4) \Gamma_4(v_5) \Gamma_4(v_6) \Gamma_4(v_8)  }{\Gamma_4(v_1) \Gamma_4(v_2) \Gamma_4(v_3) \Gamma_4(v_7) }\left( \frac{\pi}{\delta} \right)^{\frac{dim}{2}}
\end{eqnarray}
where $v_i=w_i|_{\kappa =0}$. $v_i$ will be used in next paragraph, so we explicitly write them below:  
\begin{eqnarray}\label{vs}
v_1&=& 2t-2\alpha,\\
v_2&=& t+\gamma,\\
v_3&=& t+\beta,\\
v_4&=& 3t,\\
v_5&=& 2t+2\beta+\gamma,\\
v_6&=& 2t+\beta+2\gamma,\\
v_7&=& 5t-\alpha,    \\ \label{vs2}
v_8&=& -\alpha,
\end{eqnarray}
 
According to  (\ref{vol}), this formula  essentially  gives representation of   invariant volume of simple Lie groups through Barnes' quadruple function.

In total partition function explicit $\delta \text{ and } dim$ cancel: 
 \begin{eqnarray}\label{tpfgn}
Z=Z_1Z_2= \frac{\Gamma_4(w_1) \Gamma_4(w_2) \Gamma_4(w_3) \Gamma_4(w_7)  }{\Gamma_4(w_4) \Gamma_4(w_5) \Gamma_4(w_6) \Gamma_4(w_8) }
  \frac{\Gamma_4(v_4) \Gamma_4(v_5) \Gamma_4(v_6) \Gamma_4(v_8)  }{\Gamma_4(v_1) \Gamma_4(v_2) \Gamma_4(v_3) \Gamma_4(v_7) }
 \end{eqnarray}

Let's comment on a projective invariance of partition function(s), e.g. (\ref{F2P}). The combination of quadruple functions in (\ref{F2P}) is equal to (\ref{F2}), so is projective invariant, but each function  isn't. One can ask on most general form of projective-invariant combinations of quadruple functions. Answer is more or less clear from the above - we should construct a combination of main terms such that integral is convergent, then all additional terms will cancel and it will be invariant w.r.t. the multiplication of all arguments and parameters on an arbitrary positive  number. 

\section{Level-rank duality of $SU(N)$ partition function from Barnes' functions recurrent relations}\label{levelrank}

One of the immediate consequences of representation of partition function through Barnes' multiple gamma functions is level-rank duality  for $SU(N)$ group \cite{NS,CLZ}. It is essentially (almost) invariance of the partition function under interchange of level $k$ (value of $\kappa$ in minimal normalization) and $N$. 

In our representation this symmetry appears to be a consequence of recurrent relations \cite{Barnes3,Rui} on Barnes' multiple functions. 
 These relations have an origin just in  definition (\ref{zN}): if $w=w_0+a_i, \Re w_0>0$, then sum over $n_i$ is effectively starting from $n_i=1$, with zeta function argument being $w_0$. It remains to add and subtract contribution of $n_i=0$ to get a relation: 

\begin{eqnarray}\label{rec}
\zeta_N(w_0+a_i,s|a_1,a_2,...,a_N)=\zeta_N(w_0,s|a_1,a_2,...,a_N)-\\ \nonumber
\zeta_{N-1}(w_0,s|a_1,..., a_{i-1},a_{i+1}...,a_N)
\end{eqnarray}

This straightforwardly translates into recurrence relation on multiple gamma-functions:

\begin{eqnarray}\label{recgam}
\Gamma_N(w_0+a_i,s|a_1,a_2,...,a_N)=\\
\Gamma_N(w_0,s|a_1,a_2,...,a_N)/\Gamma_{N-1}(w_0,s|a_1,..., a_{i-1},a_{i+1}...,a_N)
\end{eqnarray}

Let's apply these relations to partition function for $SU(N)$. 

Perturbative part $Z_2$ of partition function is (\ref{pfgn}). All gamma functions have parameters (2,2,N,k+N).   With the use of recurrence relation (\ref{rec}) for parameter N  specific ratios become

\begin{eqnarray}\label{pfgn1}
 \frac{\Gamma_4(w_1)}{ \Gamma_4(w_5) }&=& \Gamma_3(2(k+N)+4|2,2,2(k+N)) \\
  \frac{\Gamma_4(w_2)}{ \Gamma_4(w_4) } &=& \Gamma_3(2(k+N)|2,2,2(k+N))\\
  \frac{\Gamma_4(w_3)}{ \Gamma_4(w_8) } &=& \frac{1}{\Gamma_3(2k+2|2,2,2(k+N))}\\
    \frac{\Gamma_4(w_7)}{ \Gamma_4(w_6) } &=& \frac{1}{\Gamma_3(2(k+N)+2+2N|2,2,2(k+N))}\\ &=& \frac{\Gamma_2(2N+2|2,2)}{\Gamma_3(2N+2|2,2,2(k+N))}
\end{eqnarray}
so perturbative partition function is 

\begin{eqnarray}\label{pfgn2}
Z_2= \frac{\Gamma_3(2(k+N)+4) \Gamma_3(2(k+N))\Gamma_2(2N+2|2,2) }{\Gamma_3(2k+2)\Gamma_3(2N+2) }\left( \frac{k+N}{\pi} \right)^{\frac{dim}{2}}
\end{eqnarray}
which already seems almost symmetric on k and N, but note "dangerous" $N$ in $dim=N^2-1$. From (\ref{pfgn2}) we get nonperturbative partition function $Z_1$ by (\ref{F1}): 

\begin{eqnarray}\label{pfgn3}
Z_1 &=& \frac{\Gamma_3(2)\Gamma_3(2N+2)  }{    \Gamma_3(2N+4) \Gamma_3(2N)\Gamma_2(2N+2|2,2) }\left( \frac{\pi}{k+N} \right)^{\frac{dim}{2}} \\
&=& \frac{\Gamma_2(4|2,2)\Gamma_2(2|2,2N)  }{     \Gamma_2(2N|2,2N)\Gamma_2(2N+2|2,2) }\left( \frac{\pi}{k+N} \right)^{\frac{dim}{2}}\\
&=& \frac{\Gamma_2(4|2,2)\Gamma_1(2|2)  }{     \Gamma_2(2N+2|2,2) \Gamma_1(2N|2N)}\left( \frac{\pi}{k+N} \right)^{\frac{dim}{2}}\\
&=& \frac{\Gamma_2(4|2,2)  }{     \Gamma_2(2N+2|2,2)} \frac{1}{ \sqrt{N}}\left( \frac{\pi}{k+N} \right)^{\frac{dim}{2}}
\end{eqnarray}
where on last stage we use (\ref{gamma1}).

Altogether, partition function $Z(N,k)$ (explicitly noting its dependence on $k$ and $N$) is:

\begin{eqnarray}\label{pfgn4}
Z(N,k)=Z_1Z_2= \frac{\Gamma_3(2(k+N)+4) \Gamma_3(2(k+N)) \Gamma_2(4|2,2)}{\Gamma_3(2k+2)\Gamma_3(2N+2) } \frac{1}{ \sqrt{N}}
\end{eqnarray}
Remind that parameters of triple gamma functions are $(2,2,2(k+N))$. Now evidently 
\begin{eqnarray}\label{pfgn5}
\frac{Z(N,k)}{Z(k,N)}= \frac{\sqrt{k}}{\sqrt{N}}
\end{eqnarray}
which is exactly the statement of level-rank duality of partition function of $SU(N)$ Chern-Simons on 3d sphere \cite{NS,CLZ}.

Note that usual proofs of level-rank duality are working for physical values of N and k, particularly they have to be integers, while here we show duality of analytically extended partition function for non-integer values, also. 

\section{Analytic continuations of partition function  and their parameter's permutations} \label{reflect}

Let's consider perturbative partition function $F_2$ (\ref{F2}) at general complex values of all variables. Integrand requires   $\delta\neq 0 $, so let's put $\delta=1$ by projective transformation. Then it is evident, that one should have $\Re\alpha\neq 0, \Re\beta\neq 0, \Re\gamma\neq 0$ since otherwise there are non-integrable singularities at the poles coming from the zeros of one of $\sinh$ in denominator. This restriction divides the space of parameters into disjoint regions. Inside that regions integral can converge or diverge at large $x$ depending on values of parameters.

Function $F_2$ is invariant w.r.t. the permutations of parameters, but it is not an analytic function of parameters, as discussed in Introduction \footnote{In some paper(s) similar functions (those in \ref{vol}) are incorrectly declared to be an analytic functions on $CP^2$}.

Let's introduce notation $K_{\pm\pm\pm}(\alpha,\beta,\gamma)$ for analytic functions, which are equal to $F_2$ (\ref{F2}) in the region where signs of real parts of parameters $\alpha,\beta,\gamma$ coincide respectively with their subscripts (index). By argument given in Introduction, these functions are symmetric w.r.t. the transposition of arguments corresponding to the same signs in index, since we can interchange them smoothly by paths in the region of definition of integral. For example 

\begin{eqnarray}\label{symm}
K_{--+}(\alpha,\beta,\gamma)=K_{--+}(\beta,\alpha,\gamma)\\
\Re \alpha<0, \Re\beta<0, \Re\gamma>0   \nonumber
\end{eqnarray}
but in general $K_{--+}(\alpha,\beta,\gamma)$ is not symmetric w.r.t. the transposition of $ \beta<0$ and  $\gamma>0$. 

From definition we get relations:

\begin{eqnarray}\label{Kpmm}
K_{--+}(\alpha,\beta,\gamma)=K_{-+-}(\alpha,\gamma,\beta)=K_{+--}(\gamma,\alpha,\beta)\\
\Re \alpha<0, \Re\beta<0, \Re\gamma>0  \nonumber
\end{eqnarray}
and 
\begin{eqnarray}\label{Kppm}
K_{++-}(\alpha,\beta,\gamma)=K_{+-+}(\alpha,\gamma,\beta)=K_{-++}(\gamma,\alpha,\beta)\\
\Re \alpha>0, \Re\beta>0, \Re\gamma<0  \nonumber
\end{eqnarray}
and similar ones with arguments with same signs transposed. 

One can try to analytically continue these functions to other regions, where they don't necessarily coincide with functions K originated from that region, and calculate their difference. All relations of type (\ref{symm}),(\ref{Kpmm}) and (\ref{Kppm}) presumably will be maintained by these analytic continuations. So, for example, we can take $K_{-++}(\alpha,\beta,\gamma)$, where $\Re \alpha<0, \Re\beta>0, \Re\gamma>0$, analytically continue it to other region of arguments, e.g. $\Re \alpha>0, \Re\beta>0, \Re\gamma>0$ and  calculate difference  $K_{-++}(\alpha,\beta,\gamma) - K_{+++}(\alpha,\beta,\gamma)$. Next we can imagine to carry on analytic continuation w.r.t. the parameter $\beta $ and calculate difference between obtained function and function originated from that new region: 

\begin{eqnarray}
K_{-++}(\alpha,\beta,\gamma) - K_{+-+}(\alpha,\beta,\gamma)=\\
\left( K_{-++}(\alpha,\beta,\gamma) - K_{+++}(\alpha,\beta,\gamma)\right) +\\  \left( K_{+++}(\alpha,\beta,\gamma) - K_{+-+}(\alpha,\beta,\gamma)\right) = \\
K_{-++}(\alpha,\beta,\gamma) - K_{-++}(\beta,\alpha,\gamma)\\
\Re \alpha>0, \Re\beta<0, \Re\gamma>0
\end{eqnarray}
where at last step we use (\ref{Kpmm}). I.e. carrying on analytic continuations twice, w.r.t. the arguments with different signs of real part, and summing corresponding two relations, we  obtain a behavior of function $K$ under transposition of parameters with different signs of their real parts. 

All this seems to fit into a fiber bundle  over Vogel's plane with group of permutations as a structure group and  functions $K$ being a section(s).  

An example of such a procedure can be given for an $SU(N)$ values of parameters, which are $(\alpha/t,\beta/t,\gamma/t)=(-2/N,2/N,1)$, and we also put $\delta=t=N$ (i.e. consider volume function integral), to make a comparison with known cases. In this case job can be done in one step, since the change of sign of N corresponds to the interchange of parameters $\alpha$ and $\beta$.

Generally, for  volume function (\ref{vol}), i.e. for the case $\delta=t$, it is easy to establish that  integral converges when parameters  $\Re \alpha, \Re \beta, \Re \gamma, (\alpha+\beta+\gamma=1)$ are of different signs, and diverges otherwise (i.e. when they all are positive). On the plane $(\Re\alpha, \Re\beta)$ line $\Re\gamma=0$ corresponds to the line $\Re\alpha+\Re\beta=1$. So, lines of zero real parts of parameters divide  $(\Re\alpha, \Re\beta)$ plane on 7 regions. Similarly hyperplanes  $\Re\alpha=0, \Re\beta=0$ and  $\Re\gamma=0$ divide projective space of $\alpha, \beta, \gamma$ (i.e. $CP^2$) into seven disconnected pieces. Integral doesn't converge in one region  only, namely in the region where all real parts of parameters are positive.

Integral for $SU(N)$ is (\ref{F1}):

\begin{eqnarray}\label{suvol}
K_{-++}(-\frac{2}{N},\frac{2}{N},1)=\int^{\infty}_0 \frac{dx}{x}\left(\frac{1-e^{-x}}{4\sinh^2(\frac{x}{2N})}-\frac{N^2}{e^x-1}\right)\\
\Re N > 0
\end{eqnarray}

We are interested in change of (\ref{suvol}) under change of sign of $N$.   Let we have $N$ with $\Re N >0$. Then $N$-dependent poles of integrand of \ref{suvol} are in the points $x= i \pi k/N, k= 1,  2, ...$. Now let's move $N$ to $-N$, e.g. by multiplying on phase factor, changing from 1 to -1 in counterclockwise direction.  Then poles will move in clockwise direction and those with $k>0$ will touch the integration line $[0,\infty)$.   

When poles reach integration contour from upper semiplane and continue to move to lower semiplane, we deform contour to prevent appearance of singularity. One can imagine that deformation as a creation of a narrow sprout of the contour, which go from the real positive line to a pole (which is in the lower semiplane), turn around him in counterclockwise direction, and return to real positive line. Moving parameter $N$ to its new value, and simultaneously deforming the contour, we get a value of initial function, analytically continued to new value of parameter $N$.

Finally, when $N$ becomes $-N$, we get new contour of integration which evidently can be replaced   by line  from 0 to infinity plus small circles, enclosing poles at points  $x=- i \pi k/N, k= 1,  2, ...$ in counterclockwise direction.  Integral over line is an initial integral with $-N$ instead of $N$, which is the same. So, the value of analytically continued function on point $-N$ is equal to its value at $N$  plus $2\pi i$ times residues at poles. Residue at the pole at $x= - i \pi k/N$ is:

\begin{eqnarray}
Res_{x= -\frac{i \pi k}{N}} \left( \frac{1}{x}\left(\frac{1-e^{-x}}{4\sinh^2(\frac{x}{2N})}-\frac{N^2}{e^x-1}\right) \right)=\\ 
\frac{1+e^{2k i \pi  N} (-1+2k i \pi  N)}{4k^2 \pi ^2}
\end{eqnarray}

so we have

\begin{eqnarray} \label{permSU}
K_{-++}(-\frac{2}{N},\frac{2}{N},1)-K_{-++}(\frac{2}{N},-\frac{2}{N},1)=\\  \nonumber
-2\pi i \sum_{k=1}^{\infty}\frac{1+e^{2k i \pi  N} (-1+2k i \pi  N)}{4k^2 \pi ^2}
\end{eqnarray}

Now let's use this answer with expression \cite{Mkr3} for Barnes' G-function  throw integral (\ref{suvol}), to make contact with known reflection formula for G. This expression appears from comparison of two  calculations of  nonperturbative part of $SU(N)$ Chern-Simons theory - either directly \cite{Pe}, or through universal approach \cite{Mkr3}:

\begin{eqnarray} \label{GN}
\ln (G(1+N))=\frac{1}{2}N^2\ln N -\frac{1}{2}(N^2-N)\ln(2\pi)+ K_{-++}(-\frac{2}{N},\frac{2}{N},1)
\end{eqnarray}

From this equation, applying the procedure of sign changing of $N$ by counterclockwise rotation, and using (\ref{permSU}), we get reflection relation for Barnes' G-function:

\begin{eqnarray} \label{GNGN}
\ln \frac{G(1+N)}{G(1-N)}=\frac{i\pi}{2}N^2 + N\ln(2\pi)- i \sum_{k=1}^{\infty}\frac{1+e^{2k i \pi  t} (-1+2k i \pi  t)}{2k^2 \pi}
\end{eqnarray}
provided we choose appropriate branch of $\ln N$.

We would like to compare this with  Kinkelin's functional equation (\ref{Kin0}), in a form given in \cite{Adamchik,Adamchik2}:

\begin{eqnarray} \label{Kin}
\ln \frac{G(1+N)}{G(1-N)}=  \frac{i}{2\pi}Li_2(e^{2\pi i N})+N\ln\left( \frac{\pi}{\sin\pi N}\right) -\frac{\pi i}{2}B_2(N)
\end{eqnarray}
where $Li_2$ is a dilogarithm function, $B_2(z)=z^2-z+1/6$ - second Bernoulli polynomial. Equivalence of two forms of Kinkelin's equation can be established directly, by integration in (\ref{Kin0}), indefinite integral is \cite{WOI}: 

\begin{eqnarray}\label{Kin2}
 \int dx \, \pi x \, cot(\pi x)=x \ln(1-e^{2\pi i x})-\frac{i}{2\pi}\left( \pi^2 x^2+Li_2(e^{2\pi i x}) \right) 
\end{eqnarray}

Writing functions in the r.h.s. of (\ref{Kin}) or (\ref{Kin2}), (\ref{Kin0}) as a sums over powers of $e^{2\pi i N}$:

\begin{eqnarray}
Li_2(e^{2\pi i N})=\sum_{k=1}^{\infty} \frac{e^{2\pi i k N}}{k^2}  \\
N \ln\frac{\pi}{\sin \pi N}=N\ln 2\pi - \frac{i\pi}{2}N+i\pi N^2+N \sum_{k=1}^{\infty}\frac{e^{2\pi i k N}}{k}
\end{eqnarray}
we get:
\begin{eqnarray}
\ln \frac{G(1+N)}{G(1-N)}=  \frac{i}{2\pi}\sum_{k=1}^{\infty}\frac {e^{2\pi i N}}{k^2}+N \sum_{k=1}^{\infty}\frac {e^{2\pi i N}}{k} +\frac{i\pi}{2}N^2 +N \ln 2\pi- \frac{i\pi}{12}
\end{eqnarray}
which coincides with (\ref{GNGN}), taking into account that

\begin{eqnarray}
\sum_{k=0}^{\infty}\frac {1}{k^2}=\frac{\pi^2}{6}
\end{eqnarray}

The similar calculation we carry on for general case of arbitrary parameters. The answer seemingly gives reflection relation, which  reproduce (\ref{Kin0}) in particular case of $SU(N)$, but because in the process  one passes through divergent series, it  requires  further careful study. 

Another way of handling an analytical continuation is an application of the theory of multiple Barnes' functions \cite{Barnes3} to our representation of $F_2 $ through Barnes' gamma functions. Particularly in \cite{Barnes3} there is defined a region of parameters where sums \ref{zN} are convergent. One have to avoid an existence of accumulation points of denominators, it is achieved and sums are convergent, at general values of argument, iff parameters in complex plane all are on one side of some line crossing the origin (\cite{Barnes3}, p.387). Gamma functions in \ref{pfgn} have parameters $(-\alpha,\beta,\gamma,2\delta)$ which all are positive in physical region, so are on one side of e.g. purely imaginary numbers axis. From the other side, for other values of parameters, e.g. for all positive $(\alpha,\beta,\gamma,\delta)$ we shall have another formula for integral $F_2$, again as a combination of gamma functions with parameters $(\alpha,\beta,\gamma,2\delta)$, all positive ones. The multiple gamma functions at this two sets of parameters are presumably connected by analytical continuation and some kind of reflection formula can express the value of one through another. This requires a developed theory of Barnes' multiple gamma functions. 

Evidently,  investigation of analytic properties of functions $K_{\pm\pm\pm}$ requires (and worth) much more efforts. The modest aim of this Section is to note a problem and  establish some framework for such an investigation.

\section{Conclusion}\label{concl}
In present paper we explore recently discovered integral representation of partition function of Chern-Simons theory on 3d sphere. Particularly, we transform it into combination of quadruple Barnes' gamma functions. Presumably, this will help in study of exact properties of partition function at  finite values of parameters (coupling constant and universal parameters), due to the known global features (such as zeros and poles) of gamma functions. We assume that expression (\ref{tpfgn}) catch some fundamental features of partition function of Chern-Simons theory, so on some other 3d manifolds  partition functions of Chern-Simons  can be expressed through ratio of products of quadruple gamma function, also. 

Very probably, partition function of refined Chern-Simons theory, for which recently a universal representation is obtained \cite{KS}, can be transformed into gamma functions representation, also.

One of the immediate consequences of this representation is an extension in Section \ref{levelrank} of level-rank duality of $SU(N)$ partition function to all, particularly non-integer, values of N and level k. The proof uses recurrent relations for Barnes' multiple gamma functions. These relations exist for an arbitrary values of parameters and argument of these gamma function, so one may try to extend level-rank duality to more general  values of universal parameters, out of domain of classical algebras. 

For classical algebras one still have to check some known facts on this language of gamma functions: level-rank duality of SO and Sp algebras, $N\rightarrow -N$ duality \cite{MV1,Mkr} of SO/Sp theories, Ooguri-Vafa \cite{OV} relation between volumes of classical compact groups, some features in Sinha-Vafa \cite{SV} (see also\cite{MV1}) duality of  SO/Sp Chern-Simons theory to  topological strings theory with nonorientable surfaces included, etc. 
 
Barnes' multiple function are a kind of number theory functions, generalizations of most famous representative of this class of functions - Riemann's zeta function. Formula (\ref{vol}) express invariant volume of simple Lie groups through Barnes' quadruple gamma function, which seems to be beautiful and unexpected relation between number theory  and intrinsic geometrical properties of simple Lie algebras and groups. It is relevant here to recall previously  obtained relation \cite{Mkr2}  between classification of simple Lie algebras and certain Diophantine equations. This relation is based on an expression (\ref{gene}) of universal character on a special line. For simple Lie algebras this expression is a finite sum of exponents, i.e. it is regular on a finite complex x plane. One can ask for a solution of an inverse, in some sense, problem - at what values of universal parameters this function is regular in finite x plane? It appears that these values of parameters are in one to one correspondence with solutions of certain seven Diophantine equations on three integers $k, n$ and $m$. Each of these equations has the form $knm=ak+bn+cm$, with given integers $a,b,c$. Complete set of solutions of these equations contains, besides all simple Lie algebras, also 50 similar, in some sense, objects, among which is a $E_{7\frac{1}{2}}$ algebra \cite{LM3}.

So, we have already two interesting and non-trivial connections between universal approach to Lie algebras and gauge theories, from one side, and number theory, from the other side.

There exist another  possible direction of research which is an interpretation of some properties of number theory functions in physical language and correspondingly a possible input to number theory from physical ideas. An  example of such  an ideas  is that of proof of Riemann hypothesis on zeros of Riemann zeta-function through interpretation of them as an eigenvalues of some quantum Hermitian Hamiltonian (Polya and Hilbert, cited in \cite{Cohen}). Then its eigenvalues should be real, which hopefully can be translated into needed property of zeros to occupy  Riemann's critical straight line. Along these lines are an attempts \cite{Wolf,Julia,Knauf} to find a connection between theory  of Riemann function and the Lee-Yang theorem \cite{LY1,LY2}. So from present work's prospects the reasonable question would  be a discussion of Lee-Yang theorem for (\ref{pfgn}) with respect to universal parameters.

It is worth to mention also the questions arising from gauge/string duality, part of which is Chern-Simons theory/topological strings duality \cite{GV}.  Universal representation of partition function naturally rise question what string theory is dual to this universal partition function. There is a lot of different faces and particular cases of this question. E.g. universal form gives particularly a partition function for exceptional series of simple Lie groups, which can be used for generalization of $1/N$ expansion for exceptional groups. This will require  topological interpretation of corresponding coefficients. Note the hypothesis of Deligne \cite{Del,DM}, which from a physical perspective states that the line of exceptional simple Lie algebras is similar to that of $SU(N)$, which means that all formulae for exceptional groups (dimensions, Feynman weights, etc.) can be extended to an arbitrary point on that line (as $SU(N)$ formulae exist for all $N$, not only integer ones  \cite{H1}).  A beautiful formula of Ooguri-Vafa \cite{OV} connects $1/N$ expansion of volume of classical simple groups with virtual Euler characteristics of surfaces on $S^3$, so universal expression for volume gives rise to a question of interpretation of corresponding coefficients of expansion over universal parameters. Perhaps multiple gamma functions representation can take part in construction of these expansions. 

\section{Acknowledgments.}

Work is partially supported by Volkswagen Foundation and by the Science Committee of the Ministry of Science and Education of the Republic of Armenia under contract  13-1C232.


\begin{thebibliography}{99}
\bibitem{Mkr3}
R.L.Mkrtchyan, {\it Nonperturbative universal Chern-Simons theory}, arXiv:1302.1507 (2013), JHEP, in press.

\bibitem{W1} 
 E. Witten {\it Quantum field theory and the Jones polynomial.} Comm. Math. Phys. {\bf 121} (1989),  351-399.
 
\bibitem{V0}
P.Vogel {\it Algebraic structures on modules of diagrams.}, Preprint (1995),  
J. Pure Appl. Algebra {\bf 215} (2011), no. 6, 1292-1339.

\bibitem{V}
 P. Vogel {\it The universal Lie algebra.} Preprint (1999).

 \bibitem{LM1} 
 J.M. Landsberg, L. Manivel {\it A universal dimension formula for complex simple Lie algebras.} Adv. Math. {\bf 201} (2006), 379-407
 
\bibitem{MV1}
R.L. Mkrtchyan, A.P. Veselov, {\it Universality in Chern-Simons theory}, arXiv:1203.0766, JHEP08 (2012) 153

\bibitem{Barnes3}
E.W. Barnes, {\it On the theory of the multiple gamma function}. Trans. Cambridge Philos. Soc.
19 (1904), 374-425

\bibitem{Fad}
L. D. Faddeev, Volkov's Pentagon for the Modular Quantum Dilogarithm, arXiv:1201.6464 [math.QA], 
Functional Analysis and Applications, vol.45(4), 2011, p.65 (in Russian)

\bibitem{Rui}
S. N. M. Ruijsenaars, {\it On Barnes' Multiple Zeta and Gamma Functions}, Advances in Mathematics 156, 107-132 (2000)

\bibitem{NS}
S. G. Naculich, H. A. Riggs, and H. J. Schnitzer, Group level duality in WZW
models and Chern-Simons theory, Phys. Lett. B246 (1990) 417–422.

\bibitem{CLZ}
M. Camperi, F. Levstein, and G. Zemba, The large N limit of Chern-Simons gauge
theory, Phys. Lett. B247 (1990) 549–554.

\bibitem{Rui2}
S. N. M. Ruijsenaars, First order analytic difference equations and integrable quantum
systems, J. Math. Phys. 38 (1997), 1069-1146.

\bibitem{Kin}
Hermann Kinkelin, Ueber eine mit der gammafunction verwandte transcendente und deren anwendung auf die integralrechnung. J.Reine Angew.Math. 57 (1860), 122-158.

\bibitem{Barnes1}
E.W.Barnes, "The theory of the G-function", Quarterly Journ. Pure and Appl. Math. 31 (1900), 264–314.

\bibitem{Pe} 
 V. Periwal {\it Topological closed string interpretation of Chern-Simons theory.} Phys. Rev. Lett. {\bf 71} (1993), 1295; hep-th/9305115.

\bibitem{Adamchik}
Adamchik, Victor S., "On the Barnes Function" (2001), Computer Science Department, Paper 91,
http://repository.cmu.edu/compsci/91

\bibitem{Adamchik2}
V.S.Adamchik, Contributions to the Theory of the Barnes Function, arXiv:math/0308086 [math.CA]

\bibitem{WOI}
Wolfram online integrator, http://integrals.wolfram.com/


\bibitem{Mkr2}
R.L.Mkrtchyan, On a map of Vogel`s plane, arxiv:1209.5709.


\bibitem{LM3} J. M. Landsberg, L. Manivel  {\it The sextonions and $E_{7\frac{1}{2}}$}  Adv. Math. (2006), 201(1): 143-179, MR2204753

 
\bibitem{Cohen}
 P.B. Cohen, "Dedekind zeta functions and quantum statistical mechanics", preprint ESI 617 (1998), available at http://www.mat.univie.ac.at/~esiprpr/esi617.pdf
 
 \bibitem{Wolf}
 Marek Wolf, Applications of statistical mechanics in  number theory
 Physica A 274 (1999) 149-157
 
 \bibitem{Julia}
 B.L. Julia, "Statistical theory of numbers", from Number Theory and Physics, M. Waldschmidt, et. al. (eds.), Springer Proceedings in Physics 47 (Springer, 1989)


\bibitem{Knauf}
A. Knauf, Phases of the Number Theoretic Spin Chain, J. Stat. Phys. 73 (1993), 423-431

\bibitem{LY1}
C. N. Yang and T. D. Lee, Statistical theory of equations of state and phase transitions
I: Theory of condensation, Phys. Rev. 87, 404–409 (1952).

\bibitem{LY2}
T. D. Lee and C. N. Yang, Statistical theory of equations of state and phase transitions
II: Lattice gas and Ising model, Phys. Rev. 87, 410–419 (1952).


\bibitem{GV} R. Gopakumar, C. Vafa {\it On the Gauge Theory/Geometry Correspondence.} Adv.Theor.Math.Phys. {\bf 3} (1999) 1415-1443, arXiv:hep-th/9811131v1. 

\bibitem{OV}
H. Ooguri and C. Vafa, {\it Worldsheet derivation of a large N duality} Nucl. Phys. {\bf B 641} (2003), hep-th/0205297.

\bibitem{SV}
S. Sinha and C. Vafa, {\it SO and Sp Chern-Simons at Large N} arxive:hep-th/0012136 (2000).

\bibitem{Mkr}
R.L. Mkrtchyan {\it The equivalence of $Sp(2N)$ and $SO(-2N)$ gauge theories.} Phys. Lett. {\bf 105B} (1981), 174-176.


\bibitem{KS}
Daniel Krefl and Albert Schwarz,  {\it Refined Chern-Simons versus Vogel universality}, Journal of Geometry and Physics,  arXiv:1304.7873 [hep-th]

\bibitem{Del}
P. Deligne {\it La s\'erie exceptionnelle des groupes de Lie.} C. R. Acad. Sci. Paris, S\'erie I {\bf 322} (1996), 321-326.

\bibitem {DM}
P. Deligne and R. de Man {\it La s\'erie exceptionnelle des groupes de Lie II.} C. R. Acad. Sci. Paris, S\'erie I  {\bf 323} (1996), 577-582.  


\bibitem{H1}
G. 't Hooft {\it A planar diagram theory for strong interactions.} Nucl.Phys. {\bf B72} (1974), 461-473.





\end{thebibliography}
\end{document}